\documentclass[twocolumn,prl,superscriptaddress,showpacs,amsmath,amssymb]{revtex4-1}

\usepackage{graphicx}

\usepackage{amssymb}
\usepackage{amsmath}

\usepackage{lineno}

\usepackage{multirow}

\usepackage{url}

\begin{document}

\title{Precision Measurement of the $^6$He Half-Life and the Weak Axial Current in Nuclei}

\def\UW{Department of Physics and Center for Experimental Nuclear Physics and Astrophysics, University of Washington, Seattle, WA 98195, USA}
\def\ANL{Physics Division, Argonne National Laboratory, Argonne, IL 60439, USA}
\def\TAM{Cyclotron Institute, Texas A\&M University, College Station, TX 77843, USA}

\author{A.~Knecht} \email[]{knechta@uw.edu}\affiliation{\UW}
\author{R.~Hong}\affiliation{\UW}
\author{D.W.~Zumwalt}\affiliation{\UW}
\author{B.G.~Delbridge}\affiliation{\UW}
\author{A.~Garc\'ia}\affiliation{\UW}
\author{P.~M\"uller}\affiliation{\ANL}
\author{H.E.~Swanson}\affiliation{\UW}
\author{I.S.~Towner}\affiliation{\TAM}
\author{S.~Utsuno}\affiliation{\UW}
\author{W.~Williams}\altaffiliation{Present address: Department of Physics, Old Dominion University, Norfolk, VA 23529, USA}\affiliation{\ANL}
\author{C.~Wrede}\altaffiliation{Present address: Department of Physics and Astronomy and National Superconducting Cyclotron Laboratory, Michigan State University, East Lansing, MI 48824, USA}\affiliation{\UW}

\date{\today}

\begin{abstract}
Studies of $^6$He beta decay along with tritium can play an important role in testing \textit{ab-initio} nuclear wave-function calculations and may allow for fixing low-energy constants in effective field theories. Here, we present an improved determination of the $^6$He half-life to a relative precision of $3 \times 10^{-4}$. Our value of $806.89 \pm 0.11_\mathrm{stat}  \,^{+0.23}_{-0.19} \,_\mathrm{syst}$~ms resolves a major discrepancy between previous measurements. Calculating the statistical rate function we determined the $ft$-value to be $803.04^{+0.26}_{-0.23}$~s. The extracted Gamow-Teller matrix element agrees within a few percent with \textit{ab-initio} calculations. 
\end{abstract}



\maketitle

Precision measurements of electroweak processes in light nuclei can provide important tests of our understanding of electroweak interactions in the nuclear medium. Many interesting problems -- ranging from solar fusion to neutrino interactions and muon and pion capture processes -- depend on their correct modeling and calculation \cite{Kub10}. Recent progress in numerical techniques enable precise, \textit{ab-initio} calculations of wave functions for light nuclei to be performed with the nucleon-nucleon interaction and without assuming a frozen core of inactive particles \cite{Pie01,Bed02,Nav03}. 

The allowed weak nuclear decays driven by the axial current -- called Gamow-Teller decays -- have historically played an important role in testing wave functions because the main operator has a simple spin and isospin structure and does not possess any radial component. Systematic comparisons with shell-model wave functions showed that in order to reproduce observations the value for the weak axial coupling constant, $g_A$, had to be `quenched'. For the $sd$-shell nuclei this difference amounted to about 30\% with respect to that measured in free neutron decay \cite{Bro85,Cho93}. In addition, when charge-exchange reactions were used to explore a large fraction of the Gamow-Teller strength sum rule, evidence also showed up for `quenching of the Gamow-Teller strength' \cite{Ost92,Rap94}. However, the origin of the quenching is not completely clear. Reference~\cite{Hax00} and others \cite{Ost92} have shown that, as shell model calculations are allowed to introduce higher and higher excitations, the need to renormalize operators disappears. But it has also been pointed out that meson-exchange currents (mediating, for example, nucleon-delta excitations) could be responsible for at least some of the apparent quenching of strength \cite{Men11}.

The decays of $^3$H and of $^6$He are special because these systems are light enough that the corresponding \textit{ab-initio} calculations can be performed with precision. In particular, Refs.~\cite{Sch02,Nav03} and later \cite{Per07,Vai09} have shown that, using the case of $^3$H to fix nucleon-delta excitations, they can reproduce the $ft$-value for $^6$He to within a few percent. These two decays, then, can play an important role in testing the accuracy of nuclear wave-functions calculations \cite{Sch02,Nav03,Per07}, or as suggested in Ref.~\cite{Vai09}, in fixing low-energy constants in effective-field theory calculations \cite{Kub10}. 

\begin{figure}
  \centering
  \includegraphics[width=0.5\textwidth]{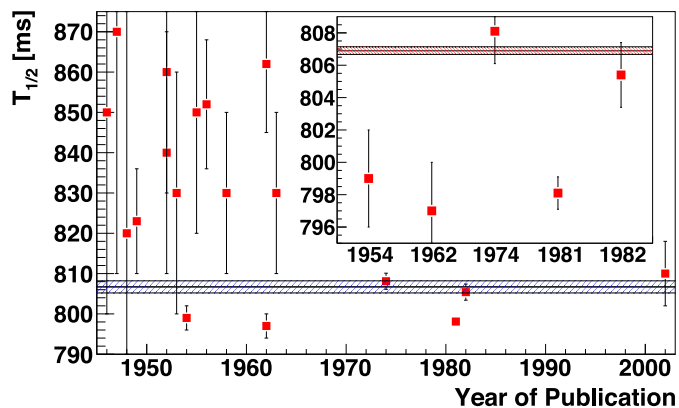}
  \caption{(Color online) Compilation of the measured $^6$He half-lives found in literature \cite{Lau66,Kli54,Bie62,Wil74,Bar81,Alb82,Ant02}. The dashed, blue band shows the half-life adopted in Ref.~\cite{Ajz84} from the average of the two values found in Ref.~\cite{Wil74,Alb82} and used in compilations ever since. The inset shows the five values with uncertainties below 1\% \cite{Kli54,Bie62,Wil74,Bar81,Alb82} with the red band depicting the value for the $^6$He half-life obtained in this paper.}
  \label{6He_Lifetime_History}
\end{figure}

In this paper we present a high-precision experimental determination of the half-life and $ft$-value for $^6$He. Except for a small branch of $\sim$$10^{-6}$ \cite{Raa09} the beta decay of $^6$He proceeds exclusively to the ground state of $^6$Li with an endpoint of 3.5~MeV. Its half-life has been previously determined by several works compiled in Fig.~\ref{6He_Lifetime_History}. 
As can be seen, the values spread over a range much wider than expected from the claimed uncertainties which makes the currently reported average and precision of $806.7 \pm 1.5$~ms \cite{Ajz84} unreliable. Averaging the five values shown in the inset in Fig.~\ref{6He_Lifetime_History} with uncertainties below 1\% and scaling the uncertainty by the square root of the $\chi^2$/dof -- as advised by the Particle Data Group in such cases \cite{Nak10} -- results in $800.6 \pm 2.0$~ms. As emphasized by the author of the last precision measurement in 1982 \cite{Alb82} this needs to be resolved by improved and higher precision experiments.

We produce $^6$He using the tandem Van de Graaff accelerator available at the Center for Experimental Nuclear Physics and Astrophysics of the University of Washington via the reaction $^7$Li($^2$H,$^3$He)$^6$He. A detailed description of the $^6$He source, which can deliver $\sim$$10^9$ atoms/s to experiments, can be found in Ref.~\cite{Kne11}. The $^6$He atoms are transferred from the source through a turbo-molecular pump to a low-background experimental area. During a period of 8~s we direct the deuteron beam onto the lithium target and the outlet of the turbo-molecular pump is connected to our measuring volume. For the following 16~s, thereby completing one cycle, we deflect the deuteron beam at the low energy end of the accelerator, close a spring loaded valve in front of the measuring volume and measure the $^6$He half-life directly by observing the decay curve.

The measuring volume consists of a 35~mm diameter, 381~mm long tube made of stainless steel sealed on one side by a 254~$\mu$m thin copper foil and on the other by a spring loaded, viton O-ring sealed valve. Randomly distributed over our measurement period of five days and taking up about half of the available data, we inserted a 19~mm diameter, 283~mm long stainless steel cylinder suspended in the center of the tube by four bolts. With the stainless steel insert we increased the wall collision frequency of $^6$He atoms by about 80\% and used that data to check for possible diffusion of the $^6$He atoms into the stainless steel surfaces. 

Directly in front of the copper foil we placed two identical, 2.5~mm thick plastic scintillators registering the betas from the decay of $^6$He. The plastic scintillators were coupled via lightguides to photomultiplier tubes. Their output signals passed through timing filter amplifiers (resulting in $\sim$100~ns long pulses) and discriminators set to the tail of the electronic noise (cutting away $\sim$1\% of the electron spectrum). We formed the coincidence of those two signals resulting in a single 25~ns long logic pulse and a total intrinsic deadtime of $\sim$130~ns, as measured on an oscilloscope. We passed this pulse through four gate generators providing signals of fixed, non-extendable deadtimes of lengths 1.9819(81)~$\mu$s, 3.9990(81)~$\mu$s, 6.0026(83)~$\mu$s, and 7.9758(83)~$\mu$s. The deadtimes were determined following the $^6$He run using the source and pulser method with the pulser sending in logic pulses at the stage of the long gate generators \cite{Bae65}. 

The four gate signals were fed into a CAMAC based scaler together with the original coincidence signal and the signals from a 1 and 100~kHz clock. The signal from the 1~kHz clock also triggered the read-out of the scaler module via the software package JAM \cite{Swa01} thereby providing 1~ms time stamps to our data stream. 

\begin{figure}
  \centering
  \includegraphics[width=0.5\textwidth]{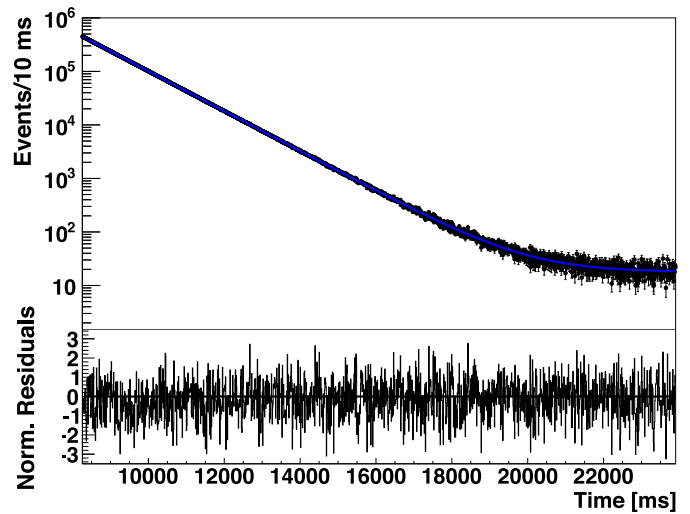}
  \caption{$^6$He decay curve for the data with initial rates $<$40~kHz and the stainless steel insert out with the corresponding residuals. The $\chi^2$/dof of the fit amounts to a typically observed 1578.2/1562.}
  \label{6He_Decay_Curve}
\end{figure}
We took data at various deuteron beam intensities to study decay-rate dependent effects. The data were grouped into five different initial rate classes: $<40$~kHz, 40 $-$ 50~kHz, 50 $-$ 60~kHz, 60 $-$ 70~kHz, 70 $-$ 80~kHz.  We corrected the deadtime losses after rebinning on a cycle-by-cycle basis by calculating the true rate $R_0$ in each time bin from the measured rate $R$ using the measured deadtimes $\tau_d$: $R_0 = R / \left( 1 - R \tau_d \right)$ \cite{Kno00}. Figure~\ref{6He_Decay_Curve} shows the decay curve for the data with initial rates $<$40~kHz and the stainless steel insert out. Also shown is a fit and the corresponding residuals with the function $R(t) = N (\exp{\left(-(t-t_0)/\tau \right)} + b)$, where the lifetime $\tau$ and the background $b$ are free parameters while $N$ is set by the total number of counts. The fit was performed using the modified $\chi^2$ method outlined in Ref.~\cite{Gri05}. We cross-checked our fit results performing both regular $\chi^2$ and maximum likelihood fits. The effects of data treatment and fitting method were studied using simulated data without noticing a particular bias.

As the systematic uncertainty due to the deadtime correction grows rapidly with rate we delayed the starting point of our fit in each of the rate groups such that for the highest rate in that group the initial rate lies at 32~kHz. Table~\ref{Tab_results} gives the final results from the averages of the different rate groups. The individual deadtime channels were combined to yield the average values for the two cases of the stainless steel insert in and out of the measuring volume. These two values are used below for an estimation of the potential diffusion of $^6$He atoms into the stainless steel surfaces.

\begin{table}
\caption{\label{Tab_results}List of the different half-lives obtained with various deadtimes and for the cases of the stainless steel insert in and out. The systematic shifts and uncertainties from Table~\ref{Tab_systematics} are not included.}
\vskip 2mm
\centering
\begin{tabular}{l c c c}
\hline
Insert & $\tau_d$ & Results [ms] & Average [ms] \\
\hline \hline
\multirow{4}{*}{out} & $\sim$2~$\mu$s & $807.01 \pm 0.11$ &  \multirow{4}{*}{$807.03 \pm 0.11$} \\
 & $\sim$4~$\mu$s & $807.03 \pm 0.11$ & \\
 & $\sim$6~$\mu$s & $807.08 \pm 0.11$ & \\
 & $\sim$8~$\mu$s & $807.02 \pm 0.11$ & \\
\hline
\multirow{4}{*}{in} & $\sim$2~$\mu$s & $807.20 \pm 0.12$ &  \multirow{4}{*}{$807.21 \pm 0.12$} \\
 & $\sim$4~$\mu$s & $807.20 \pm 0.12$ & \\
 & $\sim$6~$\mu$s & $807.21 \pm 0.12$ & \\
 & $\sim$8~$\mu$s & $807.23 \pm 0.12$ & \\
 \hline
 \hline
\end{tabular}
\end{table}
\begin{table}
\caption{\label{Tab_systematics}List of systematic shifts and uncertainties. We added the errors in quadrature to obtain the total error. Where a second value is given it corresponds to the measurements with the stainless steel insert in.}
\vskip 2mm
\centering
\begin{tabular}{l r r}
\hline
Source & Shift [ms] & Uncertainty [ms] \\
\hline \hline
Deadtime correction &  -  & 0.037 \\
$^6$He Diffusion & 0 & $^{+0.12/+0.22}_{-0}$ \\
Gain shift & -0.19 & 0.19\\
$^8$Li contamination & 0 & $^{+0}_{-0.007}$ \\
Background & 0.046 & 0.004 \\
Data correction & 0 & $0.01$\\
Deadtime drift & 0 & 0.009\\
Afterpulsing & 0 & $0.003$ \\
Pile-up & 0 & $<0.005$ \\
Clock accuracy & 0.006 &  0.011\\
\hline
Total & -0.14 & $^{+0.23}_{-0.19}$ / $^{+0.29}_{-0.19}$\\
\hline
\hline
\end{tabular}
\end{table}
In Table~\ref{Tab_systematics} we give our estimates of identified systematic shifts and uncertainties. For brevity we will only describe the most important ones in the following paragraphs. A detailed description will appear elsewhere \cite{Kne11b}.

The effect of the deadtime correction is large. The correction shifts the half-life values by $-9$~ms, $-18$~ms, $-27$~ms, and $-35$~ms for the four different deadtime channels, respectively. However, the agreement between the four values after the deadtime correction lends confidence to its validity. The uncertainty on the measured deadtimes of $\sim$8~ns translates directly into a systematic uncertainty on the half-life with its dependence extracted from the data.

Using a helium leak detector we studied the amount and time evolution of the diffusion of  helium through the walls of the measuring volume and the viton O-ring of the valve at its end. The observed amount of diffusion through the O-ring leads to a negligible shift at our level of precision. However, there most probably occurs some amount of diffusion into the walls. Any time constant $1/\tau_{loss}$ associated with this loss channel scales linearly with the wall collision frequency. From the absence of any significant difference between the two results listed in Table~\ref{Tab_results} we conclude that $\tau_{loss} \gg \tau_{^6\mathrm{He}}$ and the difference between the two results amounts to $\Delta \left( 1/\tau \right) =1/ \tau_{\mathrm{in}} - 1/\tau_{\mathrm{out}} = (-2.8 \pm 2.5) \times 10^{-7}~\mathrm{ms}^{-1} = 0.8 / \tau_{loss}$. We set the corresponding Gaussian probability density function to zero in the non-physical region \cite{Nak10} and calculate an upper limit on $1/\tau_{loss}$ at 68\% C.L. of $2\times 10^{-7}~\mathrm{ms}^{-1}$. This translates into systematic uncertainties for the insert in and out data of $^{+0.22}_{-0}$ and  $^{+0.12}_{-0}$~ms, respectively.

Examining our highest rate data we have identified traces of a small, rate-dependent shift, which is not fully accounted for by our deadtime correction. The half-lives are shifted towards higher values and we attribute this to a negative gain shift with increased rate in the photomultiplier tubes. In order to investigate its influence on our data constrained to 32~kHz, we added a parameter $k$ to our fits in order to model the effect of a linear, rate-dependent shift by substituting in our fitting function $R(t) \rightarrow R(t)\left(1-k R(t) \right)$. The resulting shift in the half-life due to including the parameter $k$ amounts to $-0.19 \pm 0.19$~ms. The fact that this parameter is compatible with no remaining rate-dependent effects was also observed when plotting the results of our half-life fits as a function of starting point of the fit window. The variations of the results were in agreement with those expected from the loss of statistics.

A potential contaminant in our system -- apart from tritium, which will not influence our measurement -- is the beta emitter $^8$Li (T$_{1/2} = 838.40(36)$~ms \cite{Fle10} with endpoint $\sim$16~MeV) produced by the reaction $^7$Li($^2$H,$^1$H)$^8$Li. The lithium atoms are not expected to reach our counting station but rather get trapped in the lithium target or in the walls during the many collisions ($\sim$$10^5$) that occur before atoms can reach the detection area. Nevertheless, in separate measurements using both our two thin detectors used for the lifetime measurements but also a thick scintillator to measure the full energy of the betas we scanned the deuteron beam energy below the reaction threshold for $^6$He production via $^7$Li($^2$H,$^3$He)$^6$He of 5.8~MeV but above the one for $^8$Li of 0.25~MeV \cite{NNDC}. While we still observed production of a beta emitter both the energy spectrum with an endpoint of 3.5~MeV and the extracted half-life of $810 \pm 5$~ms clearly identifies it as $^6$He. The observed production agrees with a rough estimate of it being produced by the reaction $^6$Li(n,$^1$H)$^6$He where the neutrons stem from $^7$Li($^2$H,n). Integrating the background-subtracted beta energy spectrum above the $^6$He endpoint and operating at our nominal deuteron beam energy we set a limit of $2 \times 10^{-4}$ at 68\% C.L. on the fraction of possible $^8$Li contamination.

Throughout the data taking we performed several background runs in which we kept the valve in front of the measuring volume closed but otherwise operated the experiment just like for the other runs. While we initially saw a significant contribution stemming from $^6$He betas penetrating through the thin-walled stainless steel bellows of our roughing pump, we were able to greatly reduce that background by shielding the bellows with lead. Combining all the background run data, we still observe a small decay structure with a half-life of 507(27)~ms and an amplitude of 6.3(3) times the value of the constant background of $0.8 \pm 0.1$~Hz. While this is most probably still coming from $^6$He that is being pumped away it could also be the result of some beam related activation. Regardless of its origin, this time-dependent background results in a systematic shift of 0.046~ms with an uncertainty of 0.004~ms.

As the result of our measurements with the stainless steel insert is dominated by the systematic uncertainty due to a potential diffusion of the $^6$He atoms into the surfaces we do not average the two values given in Table~\ref{Tab_results}. We report the data from our measurements without the insert as our final result  yielding a $^6$He half-life of  $806.89 \pm 0.11_\mathrm{stat}  \,^{+0.23}_{-0.19} \,_\mathrm{syst}$~ms. From this, we proceed to determine the $ft$-value for the beta decay of $^6$He and extract the corresponding Gamow-Teller matrix element. We calculated the $Q$-value of the decay to be $3.505208(53)$~MeV/c$^2$ using the recent $^6$He mass determination obtained in a Penning trap \cite{Bro12} and the value for $^6$Li \cite{Mou10}. This corresponds to a 4$\sigma$ shift compared to previously reported values \cite{Aud03}. The relation between the $ft$-value and the Gamow-Teller matrix element $M_\mathrm{GT}$ is
\begin{equation}
f^\star t (1+\delta ' _R) (1+\delta_\mathrm{NS}-\delta_C) = \frac{K}{G_V^2 (1 + \Delta_R^V) g_A^2 \lvert M_\mathrm{GT} \lvert ^2}
\end{equation}
following the definitions and notation of Ref.~\cite{Har09}. We set the parameters $\delta_\mathrm{NS}$ and $\delta_C$ to zero -- or equivalently absorb them into the definition of $M_\mathrm{GT}$ -- and calculate the radiative correction $\delta^\prime _R$ to be 1.0365(13)\%. We adopted the value for the parameters $K / \left(G_V^2 (1 + \Delta_R^V)\right) = 6143.62 \pm 1.66$~s from the world average of superallowed $0^+ \rightarrow 0^+$ nuclear beta decays \cite{Har09}. The statistical rate function is given by $f^\star = \int F(Z,E) p E (E-E_0)^2 f_1(E) \mathrm{d}E = f (1+\delta_s)$ where $f$ is the value of the integral in the absence of the shape-correction function $f_1(E)$ and $\delta_s$ the correction to it when including $f_1(E)$.  Here $F(Z,E)$ is the Fermi function, $p$ and $E$ the electron momentum and energy, and $E_0$ the end-point energy.  We obtain $f = 995.224(68)$ yielding an $ft$-value of $803.04^{+0.26}_{-0.23}$~s, where we added the statistical and systematic errors in quadrature. In order to take into account the shape correction we performed shell model calculations using the Cohen-Kurath interaction \cite{Coh65} and with the PWBT interactions \cite{War92} adjusted to either reproduce the experimental Gamow-Teller matrix element or the weak magnetism term, which in Holstein's notation \cite{Hol74} is $b = 68.4(7)$, determined from the width of the $0^+ \rightarrow 1^+$ transition in $^6$Li \cite{TUNL}. Both adjustments result in almost identical terms for the statistical rate function and we obtain $f^\star = 997.12(58)$.  From this we calculate the experimental value for the Gamow-Teller matrix element in $^6$He beta decay as $\lvert M_\mathrm{GT} \lvert = 2.7491(10) / \lvert g_A \lvert$.
Using $g_A = -1.2701(25)$~\cite{Nak10} determined from the decay of the free neutron, we get $\lvert M_\mathrm{GT} \lvert = 2.1645(43)$. 

Given the precision obtained in the measured half-life it is worthwhile to carefully include small contributions affecting the calculated $ft$-value. The comparisons so far \cite{Sch02,Nav03,Per07,Vai09} used an $ft$-value which dates back to Ref.~\cite{Ajz79} and a statistical rate function obtained from tabulated values in Ref.~\cite{Wil74b}. By happenstance our half-life measurement agrees with the value adopted in \cite{Ajz84}. The calculation of the statistical rate function and radiative correction together with the large shift of the $Q$-value due to the recent $^6$He mass measurements results in an experimental Gamow-Teller matrix element also in agreement with the one used in comparisons so far. However, the extraction of the experimental Gamow-Teller matrix element now stands on much more solid footing. Neglecting the error due to $g_A$ the difference between the experimental and most recent theoretical matrix element \cite{Vai09} amounts to $0.034(5)$ with theory overpredicting the experimental value by 1.5\%. The other calculations are within $\sim$5\% \cite{Sch02,Nav03,Per07}. Only \cite{Nav03} gives the calculated width of the analogous M1 transition in excellent agreement to within the experimental uncertainty of 2\%. Interesting, accompanying measurements to our result would be  improved determinations of the M1 width and of the muon capture rate on $^6$Li \cite{Deu68} to a precision of $\lesssim$1\%. The latter would directly test a possible momentum-transfer dependence of axial currents in nuclei \cite{Men11}.

In summary, we have performed the most precise measurement of the $^6$He half-life of  $806.89 \pm 0.11_\mathrm{stat}  \,^{+0.23}_{-0.19} \,_\mathrm{syst}$~ms thereby improving the precision over the currently reported value \cite{Ajz84} by a factor of 6. Our result is in good agreement with two of the previous five values \cite{Wil74,Alb82} with precisions of less than 1\% but deviates from the three others by up to 8.6$\sigma$ \cite{Kli54,Bie62,Bar81}. Because the possibility of diffusion out of the target was not directly addressed we speculate that this may be the cause of the discrepancy. Calculating the statistical rate function we determined the $ft$-value to be $803.04^{+0.26}_{-0.23}$~s. The extracted Gamow-Teller matrix element of $\lvert M_\mathrm{GT} \lvert = 2.1645(43)$ agrees within a few percent with \textit{ab-initio} calculations using the weak axial coupling constant $g_A$ measured in free neutron decay. Our precise determination allows for improved comparisons between theory and experiment and may allow using $^6$He in addition to $^3$H to fix low-energy constants in the effective-field theory description of the electroweak processes.

\begin{acknowledgments}
We thank the excellent staff at the Center for Experimental Nuclear Physics and Astrophysics. Fruitful discussions with Smarajit Triambak are gratefully acknowledged. We thank Wick Haxton and Bob Wiringa for helpful comments. This work has been supported by the US Department of Energy under DE-FG02-97ER41020. P.M. and W.W. acknowledge support by the Department of Energy, Office of Nuclear Physics, under contract DEAC02-06CH11357.
\end{acknowledgments}


\begin{thebibliography}{42}%
\makeatletter
\providecommand \@ifxundefined [1]{%
 \@ifx{#1\undefined}
}%
\providecommand \@ifnum [1]{%
 \ifnum #1\expandafter \@firstoftwo
 \else \expandafter \@secondoftwo
 \fi
}%
\providecommand \@ifx [1]{%
 \ifx #1\expandafter \@firstoftwo
 \else \expandafter \@secondoftwo
 \fi
}%
\providecommand \natexlab [1]{#1}%
\providecommand \enquote  [1]{``#1''}%
\providecommand \bibnamefont  [1]{#1}%
\providecommand \bibfnamefont [1]{#1}%
\providecommand \citenamefont [1]{#1}%
\providecommand \href@noop [0]{\@secondoftwo}%
\providecommand \href [0]{\begingroup \@sanitize@url \@href}%
\providecommand \@href[1]{\@@startlink{#1}\@@href}%
\providecommand \@@href[1]{\endgroup#1\@@endlink}%
\providecommand \@sanitize@url [0]{\catcode `\\12\catcode `\$12\catcode
  `\&12\catcode `\#12\catcode `\^12\catcode `\_12\catcode `\%12\relax}%
\providecommand \@@startlink[1]{}%
\providecommand \@@endlink[0]{}%
\providecommand \url  [0]{\begingroup\@sanitize@url \@url }%
\providecommand \@url [1]{\endgroup\@href {#1}{\urlprefix }}%
\providecommand \urlprefix  [0]{URL }%
\providecommand \Eprint [0]{\href }%
\providecommand \doibase [0]{http://dx.doi.org/}%
\providecommand \selectlanguage [0]{\@gobble}%
\providecommand \bibinfo  [0]{\@secondoftwo}%
\providecommand \bibfield  [0]{\@secondoftwo}%
\providecommand \translation [1]{[#1]}%
\providecommand \BibitemOpen [0]{}%
\providecommand \bibitemStop [0]{}%
\providecommand \bibitemNoStop [0]{.\EOS\space}%
\providecommand \EOS [0]{\spacefactor3000\relax}%
\providecommand \BibitemShut  [1]{\csname bibitem#1\endcsname}%
\let\auto@bib@innerbib\@empty
\bibitem [{\citenamefont {Kubodera}\ and\ \citenamefont {Rho}(2011)}]{Kub10}%
  \BibitemOpen
  \bibfield  {author} {\bibinfo {author} {\bibfnamefont {K.}~\bibnamefont
  {Kubodera}}\ and\ \bibinfo {author} {\bibfnamefont {M.}~\bibnamefont {Rho}},\
  }in\ \href@noop {} {\emph {\bibinfo {booktitle} {{From Nuclei to Stars,
  Festschrift in Honor of Gerald E. Brown}}}},\ \bibinfo {editor} {edited by\
  \bibinfo {editor} {\bibfnamefont {S.}~\bibnamefont {Lee}}}\ (\bibinfo
  {publisher} {World Scientific},\ \bibinfo {year} {2011})\BibitemShut
  {NoStop}%
\bibitem [{\citenamefont {Pieper}\ and\ \citenamefont {Wiringa}(2001)}]{Pie01}%
  \BibitemOpen
  \bibfield  {author} {\bibinfo {author} {\bibfnamefont {S.~C.}\ \bibnamefont
  {Pieper}}\ and\ \bibinfo {author} {\bibfnamefont {R.~B.}\ \bibnamefont
  {Wiringa}},\ }\href@noop {} {\bibfield  {journal} {\bibinfo  {journal} {Ann.
  Rev. Nucl. Part. Sci.}\ }\textbf {\bibinfo {volume} {51}},\ \bibinfo {pages}
  {53} (\bibinfo {year} {2001})}\BibitemShut {NoStop}%
\bibitem [{\citenamefont {Bedaque}\ and\ \citenamefont {van
  Kolck}(2002)}]{Bed02}%
  \BibitemOpen
  \bibfield  {author} {\bibinfo {author} {\bibfnamefont {P.~F.}\ \bibnamefont
  {Bedaque}}\ and\ \bibinfo {author} {\bibfnamefont {U.}~\bibnamefont {van
  Kolck}},\ }\href@noop {} {\bibfield  {journal} {\bibinfo  {journal} {Ann.
  Rev. Nucl. Part. Sci.}\ }\textbf {\bibinfo {volume} {52}},\ \bibinfo {pages}
  {339} (\bibinfo {year} {2002})}\BibitemShut {NoStop}%
\bibitem [{\citenamefont {Navr\'atil}\ and\ \citenamefont
  {Ormand}(2003)}]{Nav03}%
  \BibitemOpen
  \bibfield  {author} {\bibinfo {author} {\bibfnamefont {P.}~\bibnamefont
  {Navr\'atil}}\ and\ \bibinfo {author} {\bibfnamefont {W.~E.}\ \bibnamefont
  {Ormand}},\ }\href@noop {} {\bibfield  {journal} {\bibinfo  {journal} {Phys.
  Rev. C}\ }\textbf {\bibinfo {volume} {68}},\ \bibinfo {pages} {034305}
  (\bibinfo {year} {2003})}\BibitemShut {NoStop}%
\bibitem [{\citenamefont {Brown}\ and\ \citenamefont
  {Wildenthal}(1985)}]{Bro85}%
  \BibitemOpen
  \bibfield  {author} {\bibinfo {author} {\bibfnamefont {B.~A.}\ \bibnamefont
  {Brown}}\ and\ \bibinfo {author} {\bibfnamefont {B.~H.}\ \bibnamefont
  {Wildenthal}},\ }\href@noop {} {\bibfield  {journal} {\bibinfo  {journal}
  {Atom. Data Nucl. Data}\ }\textbf {\bibinfo {volume} {33}},\ \bibinfo {pages}
  {347} (\bibinfo {year} {1985})}\BibitemShut {NoStop}%
\bibitem [{\citenamefont {Chou}\ \emph {et~al.}(1993)\citenamefont {Chou},
  \citenamefont {Warburton},\ and\ \citenamefont {Brown}}]{Cho93}%
  \BibitemOpen
  \bibfield  {author} {\bibinfo {author} {\bibfnamefont {W.-T.}\ \bibnamefont
  {Chou}}, \bibinfo {author} {\bibfnamefont {E.~K.}\ \bibnamefont {Warburton}},
  \ and\ \bibinfo {author} {\bibfnamefont {B.~A.}\ \bibnamefont {Brown}},\
  }\href@noop {} {\bibfield  {journal} {\bibinfo  {journal} {Phys. Rev. C}\
  }\textbf {\bibinfo {volume} {47}},\ \bibinfo {pages} {163} (\bibinfo {year}
  {1993})}\BibitemShut {NoStop}%
\bibitem [{\citenamefont {Osterfeld}(1992)}]{Ost92}%
  \BibitemOpen
  \bibfield  {author} {\bibinfo {author} {\bibfnamefont {F.}~\bibnamefont
  {Osterfeld}},\ }\href@noop {} {\bibfield  {journal} {\bibinfo  {journal}
  {Rev. Mod. Phys.}\ }\textbf {\bibinfo {volume} {64}},\ \bibinfo {pages} {491}
  (\bibinfo {year} {1992})}\BibitemShut {NoStop}%
\bibitem [{\citenamefont {Rapaport}\ and\ \citenamefont
  {Sugarbaker}(1994)}]{Rap94}%
  \BibitemOpen
  \bibfield  {author} {\bibinfo {author} {\bibfnamefont {J.}~\bibnamefont
  {Rapaport}}\ and\ \bibinfo {author} {\bibfnamefont {E.}~\bibnamefont
  {Sugarbaker}},\ }\href@noop {} {\bibfield  {journal} {\bibinfo  {journal}
  {Ann. Rev. Nucl. Part. Sci.}\ }\textbf {\bibinfo {volume} {44}},\ \bibinfo
  {pages} {109} (\bibinfo {year} {1994})}\BibitemShut {NoStop}%
\bibitem [{\citenamefont {Haxton}\ and\ \citenamefont {Song}(2000)}]{Hax00}%
  \BibitemOpen
  \bibfield  {author} {\bibinfo {author} {\bibfnamefont {W.~C.}\ \bibnamefont
  {Haxton}}\ and\ \bibinfo {author} {\bibfnamefont {C.-L.}\ \bibnamefont
  {Song}},\ }\href@noop {} {\bibfield  {journal} {\bibinfo  {journal} {Phys.
  Rev. Lett.}\ }\textbf {\bibinfo {volume} {84}},\ \bibinfo {pages} {5484}
  (\bibinfo {year} {2000})}\BibitemShut {NoStop}%
\bibitem [{\citenamefont {Men\'endez}\ \emph {et~al.}(2011)\citenamefont
  {Men\'endez}, \citenamefont {Gazit},\ and\ \citenamefont {Schwenk}}]{Men11}%
  \BibitemOpen
  \bibfield  {author} {\bibinfo {author} {\bibfnamefont {J.}~\bibnamefont
  {Men\'endez}}, \bibinfo {author} {\bibfnamefont {D.}~\bibnamefont {Gazit}}, \
  and\ \bibinfo {author} {\bibfnamefont {A.}~\bibnamefont {Schwenk}},\
  }\href@noop {} {\bibfield  {journal} {\bibinfo  {journal} {Phys. Rev. Lett.}\
  }\textbf {\bibinfo {volume} {107}},\ \bibinfo {pages} {062501} (\bibinfo
  {year} {2011})}\BibitemShut {NoStop}%
\bibitem [{\citenamefont {Schiavilla}\ and\ \citenamefont
  {Wiringa}(2002)}]{Sch02}%
  \BibitemOpen
  \bibfield  {author} {\bibinfo {author} {\bibfnamefont {R.}~\bibnamefont
  {Schiavilla}}\ and\ \bibinfo {author} {\bibfnamefont {R.~B.}\ \bibnamefont
  {Wiringa}},\ }\href@noop {} {\bibfield  {journal} {\bibinfo  {journal} {Phys.
  Rev. C}\ }\textbf {\bibinfo {volume} {65}},\ \bibinfo {pages} {054302}
  (\bibinfo {year} {2002})}\BibitemShut {NoStop}%
\bibitem [{\citenamefont {Pervin}\ \emph {et~al.}(2007)\citenamefont {Pervin},
  \citenamefont {Pieper},\ and\ \citenamefont {Wiringa}}]{Per07}%
  \BibitemOpen
  \bibfield  {author} {\bibinfo {author} {\bibfnamefont {M.}~\bibnamefont
  {Pervin}}, \bibinfo {author} {\bibfnamefont {S.~C.}\ \bibnamefont {Pieper}},
  \ and\ \bibinfo {author} {\bibfnamefont {R.~B.}\ \bibnamefont {Wiringa}},\
  }\href@noop {} {\bibfield  {journal} {\bibinfo  {journal} {Phys. Rev. C}\
  }\textbf {\bibinfo {volume} {76}},\ \bibinfo {pages} {064319} (\bibinfo
  {year} {2007})}\BibitemShut {NoStop}%
\bibitem [{\citenamefont {Vaintraub}\ \emph {et~al.}(2009)\citenamefont
  {Vaintraub}, \citenamefont {Barnea},\ and\ \citenamefont {Gazit}}]{Vai09}%
  \BibitemOpen
  \bibfield  {author} {\bibinfo {author} {\bibfnamefont {S.}~\bibnamefont
  {Vaintraub}}, \bibinfo {author} {\bibfnamefont {N.}~\bibnamefont {Barnea}}, \
  and\ \bibinfo {author} {\bibfnamefont {D.}~\bibnamefont {Gazit}},\
  }\href@noop {} {\bibfield  {journal} {\bibinfo  {journal} {Phys. Rev. C}\
  }\textbf {\bibinfo {volume} {79}},\ \bibinfo {pages} {065501} (\bibinfo
  {year} {2009})}\BibitemShut {NoStop}%
\bibitem [{\citenamefont {Raabe}\ \emph {et~al.}(2009)\citenamefont {Raabe}
  \emph {et~al.}}]{Raa09}%
  \BibitemOpen
  \bibfield  {author} {\bibinfo {author} {\bibfnamefont {R.}~\bibnamefont
  {Raabe}} \emph {et~al.},\ }\href@noop {} {\bibfield  {journal} {\bibinfo
  {journal} {Phys. Rev. C}\ }\textbf {\bibinfo {volume} {80}},\ \bibinfo
  {pages} {054307} (\bibinfo {year} {2009})}\BibitemShut {NoStop}%
\bibitem [{\citenamefont {Lauritsen}\ and\ \citenamefont
  {Ajzenberg-Selove}(1966)}]{Lau66}%
  \BibitemOpen
  \bibfield  {author} {\bibinfo {author} {\bibfnamefont {T.}~\bibnamefont
  {Lauritsen}}\ and\ \bibinfo {author} {\bibfnamefont {F.}~\bibnamefont
  {Ajzenberg-Selove}},\ }\href@noop {} {\bibfield  {journal} {\bibinfo
  {journal} {Nuclear Physics}\ }\textbf {\bibinfo {volume} {78}},\ \bibinfo
  {pages} {1} (\bibinfo {year} {1966})},\ \bibinfo {note} {this reference
  compiles all $^6$He half-life measurements before 1966}\BibitemShut {NoStop}%
\bibitem [{\citenamefont {Kline}\ and\ \citenamefont
  {Zaffarano}(1954)}]{Kli54}%
  \BibitemOpen
  \bibfield  {author} {\bibinfo {author} {\bibfnamefont {R.~M.}\ \bibnamefont
  {Kline}}\ and\ \bibinfo {author} {\bibfnamefont {D.~J.}\ \bibnamefont
  {Zaffarano}},\ }\href@noop {} {\bibfield  {journal} {\bibinfo  {journal}
  {Phys. Rev.}\ }\textbf {\bibinfo {volume} {96}},\ \bibinfo {pages} {1620}
  (\bibinfo {year} {1954})}\BibitemShut {NoStop}%
\bibitem [{\citenamefont {Bienlein}\ and\ \citenamefont
  {Pleasonton}(1962)}]{Bie62}%
  \BibitemOpen
  \bibfield  {author} {\bibinfo {author} {\bibfnamefont {J.~K.}\ \bibnamefont
  {Bienlein}}\ and\ \bibinfo {author} {\bibfnamefont {F.}~\bibnamefont
  {Pleasonton}},\ }\href@noop {} {\bibfield  {journal} {\bibinfo  {journal}
  {Nuclear Physics}\ }\textbf {\bibinfo {volume} {37}},\ \bibinfo {pages} {529}
  (\bibinfo {year} {1962})}\BibitemShut {NoStop}%
\bibitem [{\citenamefont {Wilkinson}\ and\ \citenamefont
  {Alburger}(1974)}]{Wil74}%
  \BibitemOpen
  \bibfield  {author} {\bibinfo {author} {\bibfnamefont {D.~H.}\ \bibnamefont
  {Wilkinson}}\ and\ \bibinfo {author} {\bibfnamefont {D.~E.}\ \bibnamefont
  {Alburger}},\ }\href@noop {} {\bibfield  {journal} {\bibinfo  {journal}
  {Phys. Rev. C}\ }\textbf {\bibinfo {volume} {10}},\ \bibinfo {pages} {1993}
  (\bibinfo {year} {1974})}\BibitemShut {NoStop}%
\bibitem [{\citenamefont {Barker}\ \emph {et~al.}(1981)\citenamefont {Barker},
  \citenamefont {Ko},\ and\ \citenamefont {Scandle}}]{Bar81}%
  \BibitemOpen
  \bibfield  {author} {\bibinfo {author} {\bibfnamefont {P.~H.}\ \bibnamefont
  {Barker}}, \bibinfo {author} {\bibfnamefont {T.~B.}\ \bibnamefont {Ko}}, \
  and\ \bibinfo {author} {\bibfnamefont {M.~J.}\ \bibnamefont {Scandle}},\
  }\href@noop {} {\bibfield  {journal} {\bibinfo  {journal} {Nuclear Physics
  A}\ }\textbf {\bibinfo {volume} {372}},\ \bibinfo {pages} {45} (\bibinfo
  {year} {1981})}\BibitemShut {NoStop}%
\bibitem [{\citenamefont {Alburger}(1982)}]{Alb82}%
  \BibitemOpen
  \bibfield  {author} {\bibinfo {author} {\bibfnamefont {D.~E.}\ \bibnamefont
  {Alburger}},\ }\href@noop {} {\bibfield  {journal} {\bibinfo  {journal}
  {Phys. Rev. C}\ }\textbf {\bibinfo {volume} {26}},\ \bibinfo {pages} {252}
  (\bibinfo {year} {1982})}\BibitemShut {NoStop}%
\bibitem [{\citenamefont {Anthony}\ \emph {et~al.}(2002)\citenamefont {Anthony}
  \emph {et~al.}}]{Ant02}%
  \BibitemOpen
  \bibfield  {author} {\bibinfo {author} {\bibfnamefont {D.}~\bibnamefont
  {Anthony}} \emph {et~al.},\ }\href@noop {} {\bibfield  {journal} {\bibinfo
  {journal} {Phys. Rev. C}\ }\textbf {\bibinfo {volume} {65}},\ \bibinfo
  {pages} {034310} (\bibinfo {year} {2002})}\BibitemShut {NoStop}%
\bibitem [{\citenamefont {Ajzenberg-Selove}(1984)}]{Ajz84}%
  \BibitemOpen
  \bibfield  {author} {\bibinfo {author} {\bibfnamefont {F.}~\bibnamefont
  {Ajzenberg-Selove}},\ }\href@noop {} {\bibfield  {journal} {\bibinfo
  {journal} {Nuclear Physics A}\ }\textbf {\bibinfo {volume} {413}},\ \bibinfo
  {pages} {1} (\bibinfo {year} {1984})}\BibitemShut {NoStop}%
\bibitem [{\citenamefont {Nakamura}\ \emph {et~al.}(2010)\citenamefont
  {Nakamura} \emph {et~al.}}]{Nak10}%
  \BibitemOpen
  \bibfield  {author} {\bibinfo {author} {\bibfnamefont {K.}~\bibnamefont
  {Nakamura}} \emph {et~al.} (\bibinfo {collaboration} {Particle Data Group}),\
  }\href@noop {} {\bibfield  {journal} {\bibinfo  {journal} {J. Phys. G}\
  }\textbf {\bibinfo {volume} {37}},\ \bibinfo {pages} {075021} (\bibinfo
  {year} {2010})}\BibitemShut {NoStop}%
\bibitem [{\citenamefont {Knecht}\ \emph {et~al.}(2011)\citenamefont {Knecht}
  \emph {et~al.}}]{Kne11}%
  \BibitemOpen
  \bibfield  {author} {\bibinfo {author} {\bibfnamefont {A.}~\bibnamefont
  {Knecht}} \emph {et~al.},\ }\href@noop {} {\bibfield  {journal} {\bibinfo  {journal} {Nucl. Instrum.
  Methods A}\ }\textbf {\bibinfo {volume} {660}},\ \bibinfo {pages} {43 }
  (\bibinfo {year} {2011})}\BibitemShut {NoStop}%
\bibitem [{\citenamefont {Baerg}(1965)}]{Bae65}%
  \BibitemOpen
  \bibfield  {author} {\bibinfo {author} {\bibfnamefont {A.~P.}\ \bibnamefont
  {Baerg}},\ }\href@noop {} {\bibfield  {journal} {\bibinfo  {journal}
  {Metrologia}\ }\textbf {\bibinfo {volume} {1}},\ \bibinfo {pages} {131}
  (\bibinfo {year} {1965})}\BibitemShut {NoStop}%
\bibitem [{\citenamefont {Swartz}\ \emph {et~al.}(2001)\citenamefont {Swartz},
  \citenamefont {Visser},\ and\ \citenamefont {Baris}}]{Swa01}%
  \BibitemOpen
  \bibfield  {author} {\bibinfo {author} {\bibfnamefont {K.~B.}\ \bibnamefont
  {Swartz}}, \bibinfo {author} {\bibfnamefont {D.~W.}\ \bibnamefont {Visser}},
  \ and\ \bibinfo {author} {\bibfnamefont {J.~M.}\ \bibnamefont {Baris}},\
  }\href@noop {} {\bibfield  {journal} {\bibinfo  {journal} {Nucl. Instrum.
  Methods A}\ }\textbf {\bibinfo {volume} {463}},\ \bibinfo {pages} {354 }
  (\bibinfo {year} {2001})}\BibitemShut {NoStop}%
\bibitem [{\citenamefont {Knoll}(2000)}]{Kno00}%
  \BibitemOpen
  \bibfield  {author} {\bibinfo {author} {\bibfnamefont {G.~F.}\ \bibnamefont
  {Knoll}},\ }\href@noop {} {\emph {\bibinfo {title} {Radiation detection and
  measurement}}}\ (\bibinfo  {publisher} {New York: Wiley},\ \bibinfo {year}
  {2000})\BibitemShut {NoStop}%
\bibitem [{\citenamefont {Grinyer}\ \emph {et~al.}(2005)\citenamefont {Grinyer}
  \emph {et~al.}}]{Gri05}%
  \BibitemOpen
  \bibfield  {author} {\bibinfo {author} {\bibfnamefont {G.~F.}\ \bibnamefont
  {Grinyer}} \emph {et~al.},\ }\href@noop {} {\bibfield  {journal} {\bibinfo
  {journal} {Phys. Rev. C}\ }\textbf {\bibinfo {volume} {71}},\ \bibinfo
  {pages} {044309} (\bibinfo {year} {2005})}\BibitemShut {NoStop}%
\bibitem [{\citenamefont {Knecht}\ \emph {et~al.}()\citenamefont {Knecht} \emph
  {et~al.}}]{Kne11b}%
  \BibitemOpen
  \bibfield  {author} {\bibinfo {author} {\bibfnamefont {A.}~\bibnamefont
  {Knecht}} \emph {et~al.},\ }\href@noop {} {}\bibinfo {note} {{in
  preparation}}\BibitemShut {NoStop}%
\bibitem [{\citenamefont {Fl\'echard}\ \emph {et~al.}(2010)\citenamefont
  {Fl\'echard} \emph {et~al.}}]{Fle10}%
  \BibitemOpen
  \bibfield  {author} {\bibinfo {author} {\bibfnamefont {X.}~\bibnamefont
  {Fl\'echard}} \emph {et~al.},\ }\href@noop {} {\bibfield  {journal} {\bibinfo
   {journal} {Phys. Rev. C}\ }\textbf {\bibinfo {volume} {82}},\ \bibinfo
  {pages} {027309} (\bibinfo {year} {2010})}\BibitemShut {NoStop}%
\bibitem [{\citenamefont {{National Nuclear Data Center NNDC}}()}]{NNDC}%
  \BibitemOpen
  \bibfield  {author} {\bibinfo {author} {\bibnamefont {{National Nuclear Data
  Center NNDC}}},\ }\href@noop {} {}\bibinfo {note}
  {\url{http://www.nndc.bnl.gov}, access: October 25, 2011}\BibitemShut
  {NoStop}%
\bibitem [{\citenamefont {Brodeur}\ \emph {et~al.}(2012)\citenamefont {Brodeur}
  \emph {et~al.}}]{Bro12}%
  \BibitemOpen
  \bibfield  {author} {\bibinfo {author} {\bibfnamefont {M.}~\bibnamefont
  {Brodeur}} \emph {et~al.},\ }\href@noop {} {\bibfield  {journal} {\bibinfo  {journal} {Phys.
  Rev. Lett.}\ }\textbf {\bibinfo {volume} {108}},\ \bibinfo {pages} {052504}
  (\bibinfo {year} {2012})}\BibitemShut {NoStop}%
\bibitem [{\citenamefont {Mount}\ \emph {et~al.}(2010)\citenamefont {Mount},
  \citenamefont {Redshaw},\ and\ \citenamefont {Myers}}]{Mou10}%
  \BibitemOpen
  \bibfield  {author} {\bibinfo {author} {\bibfnamefont {B.~J.}\ \bibnamefont
  {Mount}}, \bibinfo {author} {\bibfnamefont {M.}~\bibnamefont {Redshaw}}, \
  and\ \bibinfo {author} {\bibfnamefont {E.~G.}\ \bibnamefont {Myers}},\
  }\href@noop {} {\bibfield  {journal} {\bibinfo  {journal} {Phys. Rev. A}\
  }\textbf {\bibinfo {volume} {82}},\ \bibinfo {pages} {042513} (\bibinfo
  {year} {2010})}\BibitemShut {NoStop}%
\bibitem [{\citenamefont {Audi}\ \emph {et~al.}(2003)\citenamefont {Audi},
  \citenamefont {Wapstra},\ and\ \citenamefont {Thibault}}]{Aud03}%
  \BibitemOpen
  \bibfield  {author} {\bibinfo {author} {\bibfnamefont {G.}~\bibnamefont
  {Audi}}, \bibinfo {author} {\bibfnamefont {A.}~\bibnamefont {Wapstra}}, \
  and\ \bibinfo {author} {\bibfnamefont {C.}~\bibnamefont {Thibault}},\
  }\href@noop {} {\bibfield  {journal} {\bibinfo  {journal} {Nuclear Physics
  A}\ }\textbf {\bibinfo {volume} {729}},\ \bibinfo {pages} {337} (\bibinfo
  {year} {2003})}\BibitemShut {NoStop}%
\bibitem [{\citenamefont {Hardy}\ and\ \citenamefont {Towner}(2009)}]{Har09}%
  \BibitemOpen
  \bibfield  {author} {\bibinfo {author} {\bibfnamefont {J.~C.}\ \bibnamefont
  {Hardy}}\ and\ \bibinfo {author} {\bibfnamefont {I.~S.}\ \bibnamefont
  {Towner}},\ }\href@noop {} {\bibfield  {journal} {\bibinfo  {journal} {Phys.
  Rev. C}\ }\textbf {\bibinfo {volume} {79}},\ \bibinfo {pages} {055502}
  (\bibinfo {year} {2009})}\BibitemShut {NoStop}%
\bibitem [{\citenamefont {Cohen}\ and\ \citenamefont {Kurath}(1965)}]{Coh65}%
  \BibitemOpen
  \bibfield  {author} {\bibinfo {author} {\bibfnamefont {S.}~\bibnamefont
  {Cohen}}\ and\ \bibinfo {author} {\bibfnamefont {D.}~\bibnamefont {Kurath}},\
  }\href@noop {} {\bibfield  {journal} {\bibinfo  {journal} {Nuclear Physics}\
  }\textbf {\bibinfo {volume} {73}},\ \bibinfo {pages} {1} (\bibinfo {year}
  {1965})}\BibitemShut {NoStop}%
\bibitem [{\citenamefont {Warburton}\ and\ \citenamefont
  {Brown}(1992)}]{War92}%
  \BibitemOpen
  \bibfield  {author} {\bibinfo {author} {\bibfnamefont {E.~K.}\ \bibnamefont
  {Warburton}}\ and\ \bibinfo {author} {\bibfnamefont {B.~A.}\ \bibnamefont
  {Brown}},\ }\href@noop {} {\bibfield  {journal} {\bibinfo  {journal} {Phys.
  Rev. C}\ }\textbf {\bibinfo {volume} {46}},\ \bibinfo {pages} {923} (\bibinfo
  {year} {1992})}\BibitemShut {NoStop}%
\bibitem [{\citenamefont {Holstein}(1974)}]{Hol74}%
  \BibitemOpen
  \bibfield  {author} {\bibinfo {author} {\bibfnamefont {B.~R.}\ \bibnamefont
  {Holstein}},\ }\href@noop {} {\bibfield  {journal} {\bibinfo  {journal} {Rev.
  Mod. Phys.}\ }\textbf {\bibinfo {volume} {46}},\ \bibinfo {pages} {789}
  (\bibinfo {year} {1974})}\BibitemShut {NoStop}%
\bibitem [{\citenamefont {{TUNL Nuclear Data Evaluation Project}}()}]{TUNL}%
  \BibitemOpen
  \bibfield  {author} {\bibinfo {author} {\bibnamefont {{TUNL Nuclear Data
  Evaluation Project}}},\ }\href@noop {} {}\bibinfo {note}
  {\url{http://www.tunl.duke.edu/nucldata/}, access: October 25,
  2011}\BibitemShut {NoStop}%
\bibitem [{\citenamefont {Ajzenberg-Selove}(1979)}]{Ajz79}%
  \BibitemOpen
  \bibfield  {author} {\bibinfo {author} {\bibfnamefont {F.}~\bibnamefont
  {Ajzenberg-Selove}},\ }\href@noop {} {\bibfield  {journal} {\bibinfo
  {journal} {Nuclear Physics A}\ }\textbf {\bibinfo {volume} {320}},\ \bibinfo
  {pages} {1} (\bibinfo {year} {1979})}\BibitemShut {NoStop}%
\bibitem [{\citenamefont {Wilkinson}\ and\ \citenamefont
  {Macefield}(1974)}]{Wil74b}%
  \BibitemOpen
  \bibfield  {author} {\bibinfo {author} {\bibfnamefont {D.~H.}\ \bibnamefont
  {Wilkinson}}\ and\ \bibinfo {author} {\bibfnamefont {B.~E.~F.}\ \bibnamefont
  {Macefield}},\ }\href@noop {} {\bibfield  {journal} {\bibinfo  {journal}
  {Nuclear Physics A}\ }\textbf {\bibinfo {volume} {232}},\ \bibinfo {pages}
  {58} (\bibinfo {year} {1974})}\BibitemShut {NoStop}%
\bibitem [{\citenamefont {Deutsch}\ \emph {et~al.}(1968)\citenamefont {Deutsch}
  \emph {et~al.}}]{Deu68}%
  \BibitemOpen
  \bibfield  {author} {\bibinfo {author} {\bibfnamefont {J.~P.}\ \bibnamefont
  {Deutsch}} \emph {et~al.},\ }\href@noop {} {\bibfield  {journal} {\bibinfo
  {journal} {Phys. Lett.}\ }\textbf {\bibinfo {volume} {26B}},\ \bibinfo
  {pages} {315} (\bibinfo {year} {1968})}\BibitemShut {NoStop}%
\end{thebibliography}

%

\end{document}